\documentclass[12pt]{iopart}
\usepackage{mathtext}
\usepackage{amsfonts}
\usepackage{amssymb}
\usepackage{mathrsfs}
\usepackage{bm}
\usepackage{graphicx}%

\begin{document}

\title{Exact propagators for atom-laser interactions}

\author{A. del Campo and J. G. Muga}
\eads{\mailto{qfbdeeca@ehu.es}, \mailto{jg.muga@ehu.es}}
\address{Departamento de Qu\'\i mica-F\'\i sica,
Universidad del Pa\'\i s Vasco, Apdo. 644, Bilbao, Spain}
\def\la{\langle}
\def\ra{\rangle}
\def\om{\omega}
\def\Om{\Omega}
\def\vep{\varepsilon}
\def\wh{\widehat}
\def\tr{\rm{Tr}}
\def\da{\dagger}
\newcommand{\beq}{\begin{equation}}
\newcommand{\eeq}{\end{equation}}
\newcommand{\beqa}{\begin{eqnarray}}
\newcommand{\eeqa}{\end{eqnarray}}
\begin{abstract}
A class of exact propagators describing the interaction of an $N$-level atom 
with a set of on-resonance $\delta$-lasers is obtained by means of the 
Laplace transform method. 
State-selective mirrors are described in the limit of strong lasers.
The ladder, V and $\Lambda$ configurations for a three-level atom are discussed.
For the two level case, the transient effects arising as 
result of the interaction between both a semi-infinite beam and a wavepacket with the 
on-resonance laser are examined. 

\end{abstract}


The spacetime propagator can be considered as one of the most important tools 
in quantum physics for it governs any dynamical process.
However, the knowledge of propagators corresponding to non-quadratic Hamiltonians
is severely restricted.
In this line, the spacetime propagator for a $\delta$-potential relevant to tunnelling problems
has excited much attention \cite{Bauch85,Blinder88,Crandall93,EK88,Kleber94}.
Such interactions turn out to be particularly useful to gain physical insight 
in systems where only integrated quantities are to be considered.
A thorough discussion of point interactions as solvable models using a functional 
approach can be found in \cite{Albeverio}, and 
a formalism to incorporate general point-interactions and 
dealing with different boundary conditions has been developed by Grosche \cite{Grosche,GS98}.
Even though the method is particularly suitable to calculate the energy-dependent Green function, 
a wide class of propagators was derived in such a fashion.
The incorporation of time-dependent point-interactions has been possible through different 
approaches as Duru's method \cite{Duru89} or the use of integrals of motion \cite{DMN92}.
However, most of the effort has been focused on the dynamics of structureless particles 
and to the knowledge of the authors no attention has been paid to problems involving internal levels. 
Such state of affairs contrasts dramatically with the current surge of activity in atom optics.

In this paper we use the method of Laplace transform \cite{EK88,Kleber94}
 to tackle particles with internal structure. In particular, we shall focus on exact propagators 
 for atom-laser interactions, namely, those of an atom interacting with a set of $\delta$-laser 
 on-resonance with given interatomic transitions.
The method is introduced in section \ref{2levels} 
to obtain the exact propagator for a two-level atom. 
Details of the calculations relevant to the 
following sections are here provided.  
In section \ref{3levels} the propagators for a ladder, V, and $\Lambda$ configuration (see Fig. \ref{configlaser}) 
of lasers interacting with a three-level atom are obtained. 
The general case in which a given state is coupled to an arbitrary number of levels is discussed in section \ref{Nlevels} where the high intensity limit of the laser  
is related to state selective mirrors.
Such kind of systems presents manifold applications in laser coherent control techniques such   
as cold atomic cloud compression \cite{KD99}, atom mirrors and beam splitters \cite{MS99}, and 
different schemes where fast transitions are required as in the
implementation of logic gates for ion trap quantum computing \cite{GRZC03}.
Moreover, idealised time-of-arrival measurements \cite{ML00} and 
recently proposed improvements in Ramsey-interferometry with ultracold atoms \cite{SM06} rest 
in a full quantum mechanical treatment of the dynamics of such systems.

\section{The two-level atom}\label{2levels}

In this section, we use the method of Laplace transform 
to obtain the propagator for a 
two-level atom incident on a narrow perpendicular on-resonance laser beam.
Spontaneous decay is assumed to be negligible throughout the paper and we shall consider 
effective one-dimensional systems in which the transverse momentum components 
can be neglected as it is the case for atoms in narrow waveguides \cite{LM06}.
In a laser adapted interaction picture, and using the rotating wave approximation,
the Hamiltonian describing the system is
\begin{equation}\label{h2}
{\bf H}_{\rm c} =\frac{{\widehat p}^2}{2m}\mathbf{1}_{2}+{\bf V}\delta({\widehat{x}-\xi})
=\frac{{\widehat p}^2}{2m}\mathbf{1}_{2} 
+ \frac{\hbar\Omega}{2}\delta({\widehat{x}-\xi})\,
\left(
{0 \atop 1}
{1 \atop 0}
\right), 
\end{equation}
where $\hat{p}$ is the momentum operator conjugate to $\hat{x}$, the 
ground state $|1\rangle$ is in vector-component notation ${1 \choose 0}$,
and the excited state $|2\rangle$ is ${0 \choose 1}$.
The second term in the right hand side defines the potential strength matrix ${\bf V}$ 
and $\mathbf{1}_{2}$ the two-dimensional identity matrix. 
Equation (\ref{h2}) may be regarded as the $\epsilon\rightarrow 0$ limit of a 
laser of width $\epsilon$ and Rabi frequency $\Om_L$ keeping $\Om=\Om_L \epsilon$ constant.
$\Om_L$ here and in the following is chosen to be real.
We start then by considering the free propagator for a one-channel problem on 
a Hilbert space of square integrable functions $\mathfrak{H}$ (see for instance \cite{GS98}), 
\beq
\label{freeprop}
K_{0}(x,t\vert x',0)=\sqrt{\frac{m}{2\pi i\hbar t}}
e^{\frac{im(x-x')^{2}}{2t\hbar}}.
\eeq 
In what follows we shall be interested in describing the dynamics of particles with two internal levels.
The free propagator ($\Om=0$) for states on the Hilbert space $\mathfrak{H}\otimes\mathbb{C}^{2}$ 
is given by ${\bf K}_{0}(x,t\vert x',t')=K_{0}(x,t\vert x',t')\mathbf{1}_{2}$, 
in the same interaction picture than (\ref{h2}).
Moreover, $\delta$-type of perturbations can be generally taken into account
using the method of Laplace transform \cite{Kleber94} which assumes the 
unperturbed propagator to be known.
More precisely, the full propagator can be related to the free one through 
the Lippmann-Schwinger equation \cite{EK88,Kleber94}
\beq
\fl{\bf K}(x,t\vert x',t')={\bf K}_{0}(x,t\vert x',t')
-\frac{i}{\hbar}\int_{t'}^{t}dt''\int_{-\infty}^{\infty}dx''
{\bf K}_{0}(x,t\vert x'',t''){\bf V}(x'',t''){\bf K}(x'',t''\vert x',t').
\eeq
Given that the potential has the form of a point-interaction, the integral over 
$x''$ coordinates is straightforward.
One can then take the Laplace transform with respect to $t$, which we denote with a tilde,
\fl\beq
\label{laplacels}
\fl\widetilde{{\bf K}}(x,s\vert x',0)=
\widetilde{{\bf K}}_{0}(x,s\vert x',0)
-\frac{iV_{0}}{\hbar}
\left(
{0\atop \widetilde{K}_{0}(x,s\vert\xi ,0)}
{\widetilde{K}_{0}(x,s\vert\xi ,0)\atop 0}
\right)
\widetilde{{\bf K}}(\xi,s\vert x',0),
\eeq
where we have made use of the convolution theorem 
($\mathcal{L}[\int_{0}^{t}g(t-t')f(t')dt']=\widetilde{g}(s)\widetilde{f}(s)$).
By setting $x=\xi$ it is explicitly found that 
\beq
\label{kxi}
\widetilde{{\bf K}}(\xi,s\vert x',0)=
\left(
{1\atop \frac{iV_{0}}{\hbar}\widetilde K_{0}(0,s\vert 00)}
{\frac{iV_{0}}{\hbar}\widetilde K_{0}(0,s\vert 00) \atop 1}
\right)^{-1}
\widetilde{{\bf K}}_{0}(\xi,s\vert x' ,0).
\eeq
Next, we note the exact expression for the 
Laplace transform of the single-channel free propagator (\ref{freeprop}), 
\beq
\widetilde{K}_{0}(x,s\vert x',0)=
\sqrt{\frac{m}{2i\hbar s}}e^{-\sqrt{\frac{2ms}{i\hbar}}\vert x-x'\vert},
\eeq
which becomes necessary for evaluating the inverse of the matrix.
Combining (\ref{laplacels}) and (\ref{kxi}) the Laplace 
transform of the exact full propagator is obtained, and taking 
the inverse transform one can find the spacetime propagator
\beqa
{\bf K}(x,t\vert x',0)=
{\bf K}_{0}(x,t\vert x',0)
-\frac{iV_{0}}{\hbar}\left(
{I \atop J}
{J \atop I}
\right),
\eeqa
with
\beqa
I&=&\frac{m}{2i\hbar}\mathcal{L}^{-1}\left(
\frac{e^{-\sqrt{\frac{2ms}{i\hbar}}\left(\vert x-\xi\vert+\vert\xi-x'\vert\right)}}{s-i\frac{mV_{0}}{2\hbar^{3}}}\right),\nonumber\\
J&=&\frac{m}{2i\hbar}\mathcal{L}^{-1}\left(-i\frac{V_{0}}{\hbar}\sqrt{\frac{m}{2i\hbar s}}
\frac{e^{-\sqrt{\frac{2ms}{i\hbar}}\left(\vert x-\xi\vert+\vert\xi-x'\vert\right)}}
{s-i\frac{mV_{0}}{2\hbar^{3}}}\right).
\eeqa
Fortunately in our case,
the resulting matrix element can be related after 
taking partial fractions with standard results \cite{AS65} so that
\beqa
\fl{\bf K}(x,t\vert x',t')
& = &{\bf K}_{0}(x,t\vert x',t')
-\frac{m V_{0}}{4\hbar^{2}}\sum_{\alpha=\pm}
e^{\frac{\alpha mV_{0}\left(\vert x-\xi\vert+\vert\xi-x'\vert\right)}{\hbar^{2}}
+i\frac{mV_{0}^{2}t}{2\hbar^{3}}}\nonumber\\
& & \times{\rm{erfc}}\left[\alpha\sqrt{\frac{imV_{0}^{2}t}{2\hbar^{3}}}
+\frac{1}{2}\sqrt{\frac{2m}{i\hbar t}}
\left(\vert x-\xi\vert+\vert\xi-x'\vert\right)\right]
\left(
{\alpha \atop 1}
{1 \atop \alpha}
\right).
\eeqa 
A more compact expression can be obtained rewritting the full propagator 
 in terms of the Moshinsky function (see \ref{amoshi}),
\beq
\fl{\bf K}(x,t\vert x',t')
={\bf K}_{0}(x,t\vert x',t')
-\frac{m V_{0}}{2\hbar^{2}}\sum_{\alpha=\pm}M\left(\vert x-\xi\vert+\vert\xi-x'\vert,
\alpha\kappa,\frac{\hbar t}{m}\right)
\left(
{\alpha \atop 1}
{1 \atop \alpha}
\right),
\eeq
with $\kappa=-i m V_{0}/\hbar^{2}$ for short. 
One should notice that this propagator opens up the way 
to a whole variety of problems involving quantum dynamics of two-level atoms.
A most relevant fact is that as long as the full Hamiltonian can be written as a direct sum, 
the same property holds for the propagator. Therefore, if 
${\bf H}=\bigoplus_{s=1}^{d}{\bf H}_{s}$, 
\beq
\label{dsum}
{\bf K}(x,t\vert x',t')=\bigoplus_{s=1}^{d}{\bf K}_{s}(x,t\vert x',t'),
\eeq
which becomes very useful when one wishes to study the dynamics in a given subspace.
In sections \ref{shutter} and \ref{wavepacket} we shall focus on some analytical examples dealing 
with the quantum dynamics of semi-infinite beams and wavepackets of 2-level atoms.

\section{The three level atom}\label{3levels}

\begin{figure}
\begin{center}
\includegraphics[height=3cm,angle=0]{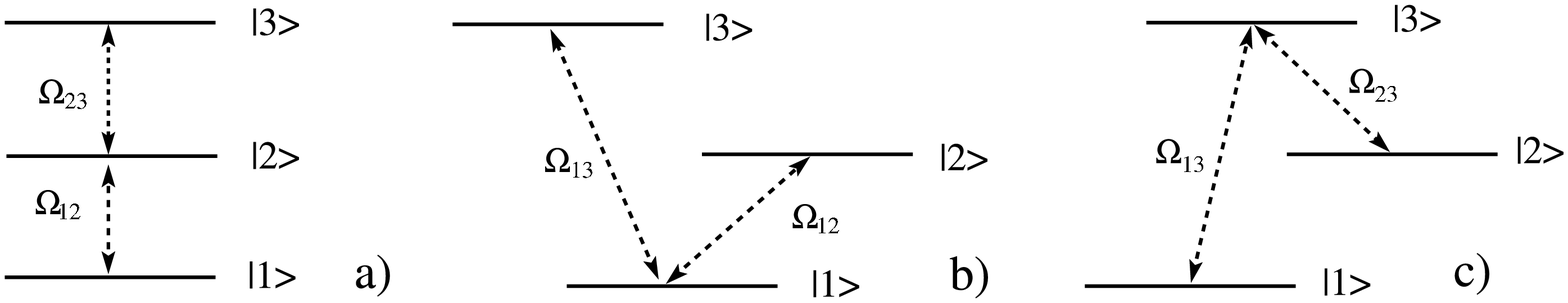}
\end{center}
\caption{\label{configlaser} 
Different configurations of a three-level atom interacting with a pair of on-resonance 
fields: a) ladder,  b) V, and c) $\Lambda$ type.
}
\end{figure}

The procedure to obtain the propagator can be extended to atoms of many-levels.
Even though the general expression can be found, the calculation is 
straightforward up to the inverse Laplace transforms.
In this section we consider a three-level atom subjected to 
two on-resonance lasers in different configurations (see Fig. \ref{configlaser}).
%
%
First, we shall focus on the ladder configuration 
which involves the transitions $|1\ra\rightarrow|2\ra$ and 
$|2\ra\rightarrow|3\ra$ by means of lasers located at the same position.
The Hamiltonian describing the system is then
\begin{equation}
\label{h3ladder}
{\bf H}_{\rm c} = \frac{{\widehat p}^2}{2m}\mathbf{1}_{3}
+{\bf V}\delta({\widehat{x}-\xi})=
\frac{{\widehat p}^2}{2m}\mathbf{1}_{3} 
+ \frac{\hbar}{2}\delta(\widehat{x}-\xi)\,
\left( \begin{array}{ccc}
0 & \Om_{12} & 0 \\
\Om_{12} & 0 & \Om_{23}\\
0 & \Om_{23} & 0 
\end{array}\right). 
\end{equation}
Following section \ref{2levels} one finds
\beqa
\fl{\bf K}(x,t\vert x',t')={\bf K}_{0}(x,t\vert x',t')
-\frac{i}{\hbar}\mathcal{L}^{-1}\left[\! 
\widetilde{{\bf K_{0}V}}(x,s\vert \xi,0)\!
\left(\mathbf{1}_{3}+\frac{i}{\hbar}\widetilde{{\bf K_{0}V}}(0,s\vert 0,0)\!\!\right)^{-1}
\!\widetilde{{\bf K}}_{0}(\xi,s\vert x',0)\right]\nonumber.
\eeqa
Plugging the explicit form of the free propagator evaluated at the different positions, 
taking partial fractions, working out the inverse Laplace transform and rewritting the 
result in terms of the Moshinsky function, one finds the exact expression for the full propagator to be   
\beqa
\fl{\bf K}(x,t\vert x',t')&=&K_{0}(x,t\vert x',t')\mathbf{1}_{3}
-\frac{m}{2\hbar^{2}}\sum_{\alpha=\pm}M\left(\vert x-\xi\vert+\vert\xi-x'\vert,
-i\alpha m\frac{\sqrt{V_{12}^{2}+V_{23}^{2}}}{\hbar^{2}},\frac{\hbar t}{m}\right)\nonumber\\
\fl& &\times\left( \begin{array}{ccc}
\frac{\alpha V_{12}^{2}}{\sqrt{V_{12}^{2}+V_{23}^{2}}} & V_{1} & \frac{\alpha V_{12}V_{23}}{\sqrt{V_{12}^{2}+V_{23}^{2}}} \\
V_{12} & \alpha\sqrt{V_{12}^{2}+V_{23}^{2}} &  V_{23}\\
\frac{\alpha V_{12}V_{23}}{\sqrt{V_{12}^{2}+V_{23}^{2}}} & V_{23} & \frac{\alpha V_{23}^{2}}{\sqrt{V_{12}^{2}+V_{23}^{2}}} 
\end{array}\right).
\eeqa

Another relevant case is the V configuration in which a pair of lasers couples the 
state $|1\ra$ with both $|2\ra$ and $|3\ra$ levels, 
in such a way that the resulting Hamiltonian is given by
\begin{equation}
\label{h3V}
{\bf H}_{\rm c} = \frac{{\widehat p}^2}{2m}\mathbf{1}_{3} 
+ \frac{\hbar}{2}\delta(\widehat{x}-\xi)\,
\left( \begin{array}{ccc}
0 & \Om_{12} & \Om_{13} \\
\Om_{12} & 0 & 0\\
\Om_{13} & 0 & 0 
\end{array}\right). 
\end{equation}
%

The exact propagator  can be found to be
\beqa
\fl {\bf K}(x,t\vert x',t')&=&K_{0}(x,t\vert x',t')\mathbf{1}_{3}
-\frac{m}{2\hbar^{2}}\sum_{\alpha=\pm}M\left(\vert x-\xi\vert+\vert\xi-x'\vert,
-i\alpha m\frac{\sqrt{V_{12}^{2}+V_{13}^{2}}}{\hbar^{2}},\frac{\hbar t}{m}\right)\nonumber\\
\fl& & \times\left( \begin{array}{ccc}
\alpha\sqrt{V_{12}^{2}+V_{13}^{2}} & V_{12} & V_{13} \\
V_{12} & \frac{\alpha V_{12}^{2}}{\sqrt{V_{12}^{2}+V_{13}^{2}}} & \frac{\alpha V_{12}V_{13}}{\sqrt{V_{12}^{2}+V_{13}^{2}}}\\
V_{13} & \frac{\alpha V_{12}V_{13}}{\sqrt{V_{12}^{2}+V_{13}^{2}}}& \frac{\alpha V_{13}^{2}}{\sqrt{V_{12}^{2}+V_{13}^{2}}} 
\end{array}\right).
\eeqa
From the mathematical point of view such configuration is actually very similar 
to $\Lambda$ scheme described by the Hamiltonian  
\begin{equation}
\label{h3L}
{\bf H}_{\rm c} = \frac{{\widehat p}^2}{2m}\mathbf{1}_{3} 
+ \frac{\hbar}{2}\delta(\widehat{x}-\xi)\,
\left( \begin{array}{ccc}
0 & 0 & \Om_{13} \\
0 & 0 & \Om_{23}\\
\Om_{13} & \Om_{23} & 0 
\end{array}\right). 
\end{equation}
whose propagator reads
\beqa
\fl {\bf K}(x,t\vert x',t')&=&K_{0}(x,t\vert x',t')\mathbf{1}_{3}
-\frac{m}{2\hbar^{2}}\sum_{\alpha=\pm}M\left(\vert x-\xi\vert+\vert\xi-x'\vert,
-i\alpha m\frac{\sqrt{V_{13}^{2}+V_{23}^{2}}}{\hbar^{2}},\frac{\hbar t}{m}\right)\nonumber\\
\fl& & \times\left( \begin{array}{ccc}
\frac{\alpha V_{13}^{2}}{\sqrt{V_{13}^{2}+V_{23}^{2}}} & \frac{\alpha V_{13}V_{23}}{\sqrt{V_{13}^{2}+V_{23}^{2}}} & V_{13} \\
\frac{\alpha V_{13}V_{23}}{\sqrt{V_{13}^{2}+V_{23}^{2}}} & \frac{\alpha V_{23}^{2}}{\sqrt{V_{13}^{2}+V_{23}^{2}}} & V_{23}\\
V_{13} & V_{23}& \alpha\sqrt{V_{13}^{2}+V_{23}^{2}} 
\end{array}\right).
\eeqa
One may then study the space-time dynamics of coherent 
trapping which results from the destructive quantum interferences
between the two transitions, for initial states in a superposition of 
the two lower levels $\vert 1\ra$ and $\vert 2\ra$ \cite{SZ97}.

\section{The $N$-level atom}\label{Nlevels}

\begin{figure}
\begin{center}
\includegraphics[height=5cm,angle=0]{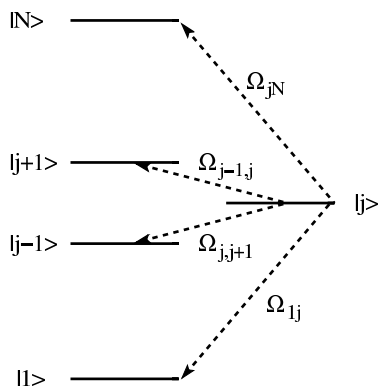}
\end{center}
\caption{\label{nlevels} 
Configuration of a $N$-level atom interacting with a $(N-1)$ on-resonance 
fields coupling the $\vert j\ra$ to any other level. 
}
\end{figure}

In this section we extend the previous approach to an 
$N$-level system living on $\mathfrak{H}\otimes\mathbb{C}^{N}$.
Suppose that an $N$-level atom is subjected to the action of 
$N$ on-resonance lasers ($\hbar\Om_{ij}/2=V_{i}$) 
all of which are located at the same position and 
couple the $\vert j\ra$ level 
($1\leq j\leq N$) with the $(N-1)$ levels as shown in Fig. (\ref{nlevels}),

%
%
\begin{equation}
\label{hN}
\fl{\bf H}_{\rm c} 
= \frac{{\widehat p}^2}{2m}\mathbf{1}_{N} 
+ \frac{\hbar}{2}\delta({\widehat{x}-\xi})\,
\left( \begin{array}{cccccccc}
0 & \cdots & 0 & \Om_{1j} & 0 & \cdots & 0 \\
\vdots & & \vdots & \vdots & \vdots & & \vdots \\
0 & \cdots & 0 & \Om_{j-1,j} & 0 & \cdots & 0 \\
\Om_{1j} & \cdots & \Om_{j-1,j} & 0 & \Om_{j+1,j} & \cdots & \Om_{N,j}\\
0 & \cdots & 0 & \Om_{j+1,j} & 0 & \cdots & 0 \\
\vdots & & \vdots & \vdots & \vdots & & \vdots \\
0 & \cdots & 0 & \Om_{Nj} & 0 & \cdots & 0 \\
\end{array}\right). 
\end{equation}
Simplified configurations can be obtained just by 
setting certain coupling elements to zero.
The exact propagator for such system reads
\beqa
\fl{\bf K}(x,t\vert x',t') = K_{0}(x,t\vert x',t')\mathbf{1}_{N}
-\frac{m V_{m}^{-1}}{2\hbar^{2}}\sum_{\alpha=\pm}M\left(\vert x-\xi\vert+\vert\xi-x'\vert,
-i\alpha m\frac{V_{m}}{\hbar^{2}},\frac{\hbar t}{m} \right)\nonumber\\
\fl \times
\left( \begin{array}{ccccccc}
\alpha V_{1}^{2} & \cdots & \alpha V_{1}V_{j-1} & V_{1}V_{m} &
\alpha V_{1}V_{j+1} & \cdots & 
\alpha V_{1}V_{N} \\

\vdots & \ddots & \vdots & \vdots & \vdots & \ddots & \vdots\\

\alpha V_{j-1}V_{1} & \cdots & \alpha V_{j-1}^{2} 
& V_{j-1}V_{m} & \alpha V_{j-1}V_{j+1}
&  \cdots & \alpha V_{j-1}V_{N}\\

V_{1}V_{m} & \cdots & V_{j-1}V_{m} & \alpha V_{m}^{2} & V_{j+1}V_{m} & \dots & V_{N}V_{m}\\

\alpha V_{j+1}V_{1}& \cdots &  \alpha V_{j+1}V_{j-1} 
& V_{j+1}V_{m} & \alpha V_{j+1}^{2}
& \cdots & \alpha V_{j+1}V_{N}\\

\vdots & \ddots & \vdots & \vdots & \vdots & \ddots & \vdots\\

\alpha V_{N}V_{1} & \cdots & \alpha V_{N}V_{j-1} & V_{N}V_{m} &
\alpha V_{N}V_{j+1} & \cdots & 
\alpha V_{N}^{2}
\end{array} \right),
\eeqa 
where $V_{m}=\sqrt{\sum_{i\neq j}V_{i}^{2}}$.
Incidentally, the ladder configuration in which only successive levels are coupled to 
each other, does not admit a similar generalisation to $N$-level given the 
complexity of the inverse matrix.

\subsection{Intense lasers and propagators for state-selective mirrors}

For the one-channel case, the effect of an infinite strength 
of the $\delta$-potential is tantamount to a hard-wall boundary condition \cite{Kleber94,AD04}.
This feature has been extensively exploited in the exact perturbation 
theory developed by Grosche to obtain a wide class of Green functions 
and propagators including boundaries 
(say, in half-space or boxes) \cite{Grosche,GS98}.  
In atom optics, it is also well-known that a very intense laser behaves 
as a totally reflective mirror for the states coupled by it.
Therefore, we shall consider the limit in which one of the lasers  
is made infinitely strong. 
In order to do so, it is convenient to note that the following relation holds 
 \cite{Kleber94}, 
\begin{displaymath}
\fl\lim_{V_{j} \to \infty}
\frac{mV_{j}}{\hbar^{2}}M\!\!\left(\!\vert x-\xi\vert+\vert\xi-x'\vert,
-i\alpha m\frac{\sqrt{\sum_{i=1}^{N}V_{i}^{2}}}{\hbar^{2}},\frac{\hbar t}{m}\right)
\!=\!\alpha K_{0}\left(\vert x-\xi\vert+\vert\xi-x'\vert,t\Big\vert 0,0\right).
\end{displaymath}
The general expression for the propagators with $N$-lasers in which the $n$-th 
one (coupling the states $\vert j\ra$ and $\vert n\ra$) is made infinitely strong becomes diagonal,
\beqa
\fl\left[{\bf K}(x,t\vert x',t')\right]_{ik}&=&K_{0}(x,t\vert x',t')\delta_{ik}
-K_{0}\left(\vert x-\xi\vert,t\Big\vert\vert x'-\xi\vert,t'\right)\delta_{ik}(\delta_{jk}+\delta_{nk}),
\eeqa
$\delta_{ij}$ being the Kr\"onecker delta.
The laser  behaves then 
as a totally reflecting mirror 
selective to the sates coupled by it, and suppresses for such states  
any possible excitation due to the presence of other lasers located 
at the same position. 

Another interesting case is the on-resonance excitation 
of a couple of levels with high-intensity lasers.
This leads to the limit where the strength coefficients of the two lasers go to infinite, 
$V_{l},V_{n}\rightarrow\infty$ and $n>l$, but its ratio is kept constant, $V_{l}/V_{n}=c$ with $c>0$.
For such case one finds
\beqa
\fl\left[{\bf K}(x,t\vert x',t')\right]_{ik}=  
-K_{0}\left(\vert x-\xi\vert,t\Big\vert\vert x'-\xi\vert,t'\right)\frac{c}{\sqrt{1+c^{2}}}\left(\delta_{ni}\delta_{lk}+\delta_{li}\delta_{nk}\right)\nonumber\\
\fl+\delta_{ik}\left[K_{0}(x,t\vert x',t')-K_{0}\left(\vert x-\xi\vert,t\Big\vert\vert x'-\xi\vert,t'\right) \left(\delta_{kj}+\frac{c^{2}}{\sqrt{1+c^{2}}}\delta_{lk}+\frac{1}{\sqrt{1+c^{2}}}\delta_{nk}\right)
\right].
\eeqa
The difference now arises from the fact that the state-selective mirrors 
have a finite reflectivity for states $\vert l\ra$ and $\vert n\ra$, 
to which excitation is allowed. However, notice that for the $\vert j\ra$ state the 
laser still mimics a totally reflecting mirror.

\section{Moshinsky shutter}\label{shutter}

\begin{figure}
\begin{center}
\includegraphics[height=6cm,angle=0]{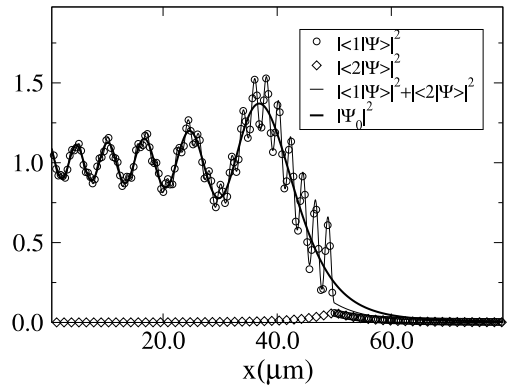}
\end{center}
\caption{\label{figdp} 
Probability density of the ground and excited states, 
total probability density in the presence of the laser and for the free case, 
$50$ ms after removing the shutter initially located at $50$ $\mu$m. 
The incident beam moves at $v=1$ cm/s (mass of $^{87} Rb$). 
The picture is taken at the instant in which the classical profile 
(the step function $\Theta(vt-x)$) reach the laser.
}
\end{figure}
%
We next study a time-dependent multi-channel scattering problem 
and consider a monochromatic beam of two-level atoms 
in its ground-state incident on a totally absorbing shutter 
which is suddenly removed at time $t=0$.
Such kind of setup is usually referred to as a Moshinsky shutter, 
ever since the seminal paper \cite{Moshinsky52} 
which led to the discovery of diffraction in time. 
The conditions on the reflectivity of the shutter can be easily modified 
to more general cases \cite{Moshinsky52,DM05}.
The initial state is then of the form 
\beq
\label{ic}
{\bf \Psi}(x,t=0)=e^{ikx}\Theta(-x)\vert 1\ra,
\eeq
where $\Theta(x)$ is the Heaviside step function.
Equation (\ref{ic}) is an obvious generalisation of the Moshinsky type 
of initial condition for a single-channel problem, 
discussed in the context of diffraction in time.
We note that such kind of state is not normalisable in the usual sense, 
yet accurately describes certain experimental setups \cite{GZ91} 
and provides a basis for wavepacket analysis.

The time evolution can be studied using the superposition principle.
If we consider first the free evolution, with $\Om=0$, 
then the result is that of diffraction in time,
\beqa
{\bf \Psi}_{0}(x,t)&=&\int_{-\infty}^{\infty}dx'{\bf K}_{0}(x,t\vert x',t'){\bf \Psi}(x,t=0)
\nonumber\\
&=& M(x,k,\hbar t/m)\vert 1\ra. 
\eeqa
This solution was found by Moshinsky and has been 
observed in a wide variety of experiments with 
ultracold neutron interferometry for the one-channel case \cite{GZ91}.
It is a well-known fact that it tends with increasing time to the stationary wavefunction. 
We further notice that due to the absence of coupling between the internal states 
in ${\bf K}_{0}(x,t\vert x',t')$, 
the excited state $\vert 2\ra$ is not populated, in agreement with (\ref{dsum}).

An alternative configuration in which the beam tunnels through a 
$\delta$-barrier was discussed in \cite{EK88,GCRV99,HGC03}.

Let us now look at the time evolution in the presence of the laser.
Using the integral (\ref{integral}) in \ref{amoshi}, 
the exact solution can be found in close-form,
\beqa
\label{eqshutter}
{\bf \Psi}(x,t)&=&\int_{-\infty}^{\infty}dx'{\bf K}(x,t\vert x',t'){\bf \Psi}(x,t=0)
\nonumber\\
&=&{\bf \Psi}_{0}(x,t)+%
\frac{1}{2}\sum_{\alpha=\pm} {\alpha \choose 1}\frac{\kappa}{k-\alpha\kappa}\nonumber\\
& & \times\bigg\{[M(\vert x-\xi\vert+\xi,k,\hbar t/m)
-M(\vert x-\xi\vert+\xi,\alpha\kappa,\hbar t/m)]\bigg\}.
\eeqa
%
\begin{figure}
\begin{center}
\includegraphics[height=6cm,angle=0]{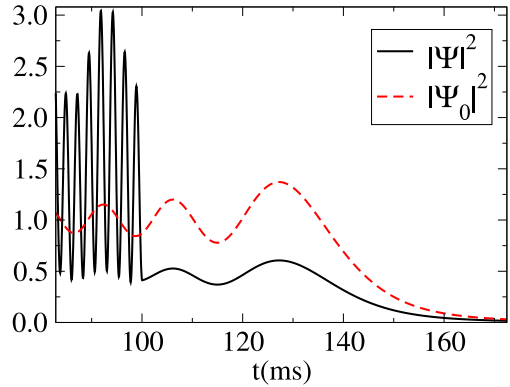}
\end{center}
\caption{\label{delay} 
Total probability density for an incident beam in the ground state 
with the laser and for free space, 
$150$ ms after removing the shutter initially located at $100$ $\mu$m. 
The incident beam moves at $v=1$ cm/s (mass of $^{87} Rb$),  
exhibiting interference for all $x<\xi$, as a result of the reflexion from the laser.}
\end{figure}
%

The total probability density  is plotted in Fig. \ref{figdp}. 
The velocity of the incident beam is chosen in all simulations to satisfy 
$v=\hbar q/m=V_{0}/\hbar$, for 
which one finds, after solving the two-channel stationary problem, that the reflexion and transmission probabilities in the excited state,  
$\vert R_2\vert^{2}=\vert T_2\vert^{2}$, are maximised and indeed equal those in the ground state, 
$\vert R_1\vert^{2}=\vert T_1\vert^{2}=1/4$. 
The paradigmatic oscillations  on the probability density, main feature of the 
diffraction in time  described by the Moshinsky 
function ($\vert{\bf\Psi}_{0}\vert^{2}$), is modified for all $x<\xi$ 
due to the interference 
which arises with the reflected part.
Indeed, for later times such interference completely dominates and 
the density profile is dramatically perturbed as shown in Fig. \ref{delay}.
%
\begin{figure}
\begin{center}
\includegraphics[height=6cm,angle=0]{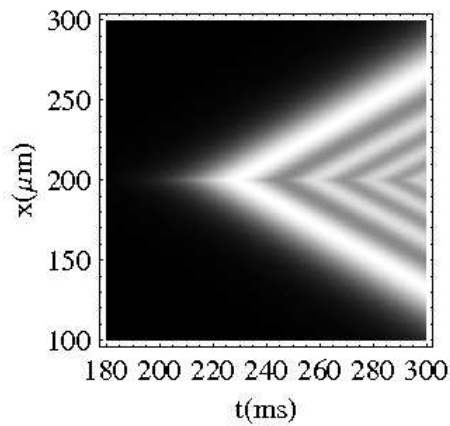}
\end{center}
\caption{\label{source2} 
Space-time density plot of the population in the excited state, 
$\vert\la 2\vert{\bf\Psi}\ra\vert^{2}$, 
with $v=1$ cm/s, and the position of the laser $\xi=200$ $\mu$m. 
The grey scale changes from dark to light as the function values increase.
}
\end{figure}

In Fig. \ref{source2} we plot the time evolution of the probability density 
in the excited state $\vert 2\ra$. 
The delta laser is shown to behave as a 
point-like source of atoms. Moreover, the pattern exhibits diffraction in 
both time and space domain.

\section{Wavepackets dynamics}\label{wavepacket} 

Many experiments deal with finite samples rather than beams.
In the one-channel case, the tunnelling dynamics of wavepackets 
through narrow barriers have been examined in a series of works \cite{EK88,AD04,GM06}.
In addition, the phenomenon of quantum deflection was predicted in the presence of 
semi-transparent and perfect mirrors \cite{DA00,DA02,AD02}.
 
\begin{figure}
\begin{center}
\includegraphics[height=6cm,angle=0]{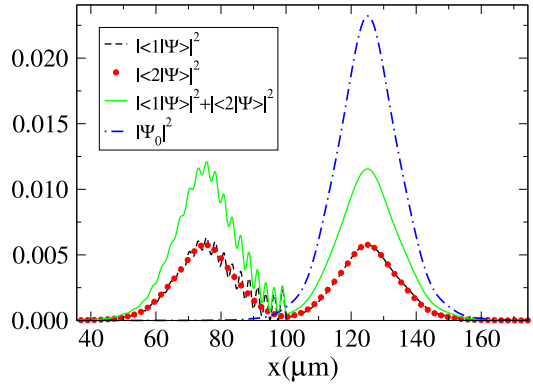}
\end{center}
\caption{\label{wpfig} 
Ground, excited and total probability densities for an incident sine-wavepacket ($n=1$)
released at $t=0$ from a hard wall trap of size $L=50\mu$m and centered at $25\mu$m from the origin,
 with $q=mV_{0}/\hbar^{2}$ 
and velocity $1$cm/s, after $100$ ms of evolution. The position of the laser is $\xi=100\mu$m. 
The freely time evolved sine wavepacket is also plotted.
}
\end{figure}

We next consider normalisable states belonging to 
$\mathcal{L}^{2}(\mathbb{R})\otimes\mathbb{C}^{2}$. 
In particular we study the dynamics of eigenstates of a hard-wall trap 
which are released at time $t=0$ and launched with momentum 
$\hbar q$ against the on-resonance delta laser. 
All-optical box traps have been recently been obtained in the laboratory 
\cite{MSHCR05}. 
More precisely we assume an initial sine-wavepacket given by
\beq
{\bf \Psi}(x,t=0)=\frac{1}{2i}\sqrt{\frac{2}{L}}
\sum_{\beta=\pm}\beta e^{iq_{n\beta}x}\chi_{[0,L]}\vert 1\ra
\eeq
with $q_{n\beta}=q+\beta n\pi/L$.
The expansion of such state has recently been discussed 
at the single-channel level in free space \cite{Godoy02}, 
in the presence of gravity \cite{DM06a}, 
and when generalised to the many-body Tonks-Girardeau regime \cite{DM05b}.
For the interaction with the on-resonance delta laser 
one can actually propagate in time this initial condition. 
The time evolved wavefunction is
\beqa
\label{phit}
\fl{\bf \Psi}_{0}(x,t)&=&\int_{-\infty}^{\infty}dx' 
K_{0}(x,t\vert x',t=0){\bf \Psi}(x',t'=0)\nonumber\\
\fl&=&\frac{1}{4i}\sqrt{\frac{2}{L}}\sum_{\beta=\pm}\beta
\big[e^{i q_{n\beta}L}M(x-L, q_{n\beta},\hbar t/m)
-M(x, q_{n\beta},\hbar t/m)\big]\vert 1\ra
\eeqa
and,
\beqa
\fl{\bf \Psi}(x,t)& = & {\bf \Psi}_{0}(x,t)+ \frac{1}{4i}\sqrt{\frac{2}{L}}\sum_{\alpha,\beta=\pm}
{\alpha \choose 1}\frac{\beta\kappa}{k-\alpha\kappa}\nonumber\\
\fl& & \times\bigg\{e^{ip_{\beta}L/\hbar}
\left[M(\vert x-\xi\vert+\xi-L,k,\hbar t/m)
-M(\vert x-\xi\vert+\xi-L,\alpha\kappa,\hbar t/m)\right]\nonumber\\
\fl& & -
\left[M(\vert x-\xi\vert+\xi,k,\hbar t/m)
-M(\vert x-\xi\vert+\xi,\alpha\kappa,\hbar t/m)\right]\bigg\}.
\eeqa
Figure \ref{wpfig} shows the dynamics when the incident momentum is such 
that maximum excitation is achieved.
%
%
The laser acts as a beam splitter dividing the wavepacket into two parts, 
the transmitted one being similar in shape to the freely evolving wavepacket, 
whereas the reflected part exhibits interference.
Notice that (\ref{eqshutter}) and (\ref{phit}) admit a simple generalisation for the 
respective $N$-level problem. 
In order to do so it suffices to consider the suitable momenta and prefactor of 
the matrix in the propagator for each of the channels.

\section{Discussion}

We have generalised the method of Laplace transform to include 
point-like perturbations in the dynamics of particles with internal structure.
In  such a fashion we have obtained the spacetime propagator for an $N$-level atom 
interacting with a set of on-resonance delta lasers.
For strong lasers, the propagators for state-selective mirrors have been obtained. 
A similar procedure could be applied to the Green function 
for which an exact perturbation theory 
has been developed and extensively discussed by Grosche \cite{GS98}.
In the 3-level atom case, the ladder,  V, and $\Lambda$ configuration have been considered.
The inclusion of time dependence in the propagators 
to adiabatically turn the lasers on and off independently from each other is 
an open problem which would provide access to the space-time dynamics 
of coherent population trapping and other relevant phenomena \cite{SZ97}.
The dynamics of a semi-infinite beam and a wavepacket of two-level atoms 
incident on the laser, with a straightforward generalisation for the $N$-level case, 
 have also been worked out. 

\ack{
This paper has benefited from inspiring comments by F. Delgado, D. Seidel, 
and I. L. Egusquiza.
This work has been supported by Ministerio de Educaci\'on y Ciencia
(BFM2003-01003) and UPV-EHU (00039.310-15968/2004).
A.C. acknowledges financial support by the Basque Government (BFI04.479). 
}

\begin{appendix}

\section{The Moshinsky function}\label{amoshi}

The Moshinsky function arises in most of the problems where 1D quantum dynamics 
involves sharp boundaries well in time or space domains.
Similarly it is found when considering free 
propagators perturbed with point-interactions.

Its standard definition reads
\beq
M(x,k,\tau):=\frac{e^{i\frac{x^{2}}{2 \tau}}}{2}w(-z),
\eeq
where
\beq
z=\frac{1+i}{2}\sqrt{\tau}\left(k-\frac{x}{\tau}\right),
\eeq
and the so called Faddeyeva function 
\cite{Faddeyeva} $w$ is explicitly defined as
\beq
w(z):= e^{-z^{2}}{\rm{erfc}}(-i z)=\frac{1}{i\pi}\int_{\Gamma_{-}}du
\frac{e^{-u^{2}}}{u-z},
\eeq
$\Gamma_{-}$ being a contour in the complex $z$-plane which goes from
$-\infty$ to $\infty$ passing below the pole.
After \cite{Moshinsky52}, $M(x,k,\tau)$ has been named 
the Moshinsky function.

For the exact time evolution in the presence of a laser with a 
cut-off plane wave or hard-wall eigenstates as initial conditions, 
we find integrals of the form
\beqa
\label{integral}
\fl\int^{x}dx'e^{ikx'}M(ax'+b,c,\tau)=\frac{e^{ikx}}{i(k+ca)}
\left[M(ax+b,c,\tau)-M(ax+b,-k/a,\tau)\right].
\eeqa
For more details we refer the reader to \cite{EK88}.

\end{appendix}


\section*{References}
\begin{thebibliography}{10}

\bibitem{Bauch85} Bauch D 1985 {\it Nuovo Cimento} B {\bf 85} 118
\bibitem{Blinder88} Blinder S M 1988 {\it Phys. Rev.} A {\bf 37} 973
\bibitem{Crandall93} Crandall R E 1993 {\it J. Phys. A: Math. Gen.} {\bf 26} 3627
\bibitem{EK88} Elberfeld W and Kleber M 1988 {\it Am. J. Phys.} {\bf 56} 154 
\bibitem{Kleber94} Kleber M 1994 {\it Phys. Rep.} {\bf 236} 331
\bibitem{Albeverio} Albeverio A, Gesztezy F, Hoegh-Krohn R J and Holden H 1988
{\it Solvable models in quantum mechanics} (Berlin: Springer)
\bibitem{Grosche} Grosche C 1993 {\it Phys. Rev. Lett.} {\bf 71} 1; 
Grosche C 1995 {\it J. Phys. A: Math. Gen.} {\bf 28} L99
\bibitem{GS98} Grosche C and Steiner F 1998 {\it Handbook of Feynman path integrals}, 
{\bf 145} {\it Springer Tracts in Modern Physics} (Berlin: Springer)
\bibitem{Duru89} Duru I H 1989 {\it J. Phys. A: Math. Gen.} {\bf 22} 4827 
\bibitem{DMN92} Dodonov V V, Man'ko V I, and Nikonov D E 1992 {\it Phys. Lett.} A {\bf 162} 359
\bibitem{KD99} Khaykovich L and Davidson N 1999 {\it J. Opt. Soc. Am.} B {\bf 16} 702
\bibitem{MS99} Metcalf H J and van der Straten P 1999 
{\it Laser cooling and trapping} (New York: Springer)
\bibitem{GRZC03} Garc\'ia-Ripoll J J, Zoller P, and Cirac J I 2003 {\it Phys. Rev. Lett.} {\bf 91} 1
\bibitem{ML00} Muga J G and Leavens C R 2000 {\it Phys. Rep.} {\bf 338} 353 
\bibitem{SM06} Seidel D and Muga J G, quant-ph/0602023
\bibitem{LM06} Lizuain I and Muga J G, quant-ph/0607015
\bibitem{AS65} Abramowitz A and Stegun I A 1965
{\it Handbook of Mathematical Functions} (New York: Dover)
\bibitem{SZ97} Scully M O and Zubairy M S 1997 {\it Quantum optics} (Cambridge: Cambridge)
\bibitem{AD04} Andreata M A and Dodonov V V 2004 {\it J. Phys. A: Math. Gen.} {\bf 37}, 2423
\bibitem{Moshinsky52} Moshinsky M 1952 {\it Phys. Rev.} {\bf 88} 625; 
Moshinsky M 1952 {\it ibid} {\bf 88} 625 
\bibitem{DM05} del Campo A and Muga J G 2005 {\it J. Phys. A: Math. Gen.} {\bf 38} 9803
\bibitem{GZ91} G\"ahler R and Zeilinger A 1991 {\it Am. J. Phys.} {\bf 59} 316 
\bibitem{GCRV99} Garc\'\i a-Calder\'on G, Rubio A, and Villavicencio J 1999 Phys. Rev. A {\bf 59} 1758
\bibitem{HGC03} Hern\'andez A and Garc\'\i a-Calder\'on G 2003 Phys. Rev. A {\bf 68} 014104
\bibitem{GM06} Granot E and Marchewka A, quant-ph/0601069
\bibitem{DA00} Dodonov V V and Andreata M A 2000 {\it Phys. Lett.} A {\bf 275} 173 
\bibitem{DA02} Dodonov V V and Andreata M A 2002 {\it Laser Phys.} {\bf 12} 57
\bibitem{AD02} Andreata M A and Dodonov V V 2002 {\it J. Phys. A: Math. Gen.} {\bf 35}, 8373
\bibitem{MSHCR05} Meyrath T P, Schreck F, Hanssen J L, Chuu C-S and Raizen M G 2005
{\it Phys. Rev.} A {\bf R71} 041604
\bibitem{Godoy02} Godoy S 2002  {\it Phys. Rev.} A {\bf 65} 042111
\bibitem{DM06a} del Campo A and Muga J G 2006 {\it J. Phys. A: Math. Gen.} {\bf 39}, 5897
\bibitem{DM05b} del Campo A and Muga J G 2006 {\it Eur. Phys. Lett.} {\bf 74}, 965
\bibitem{Faddeyeva} Faddeyeva V N and Terentev N M 1961 
{\it Mathematical Tables: Tables of the values of 
the function $w(z)$ for complex argument} (New York: Pergamon) 
 


\end{thebibliography}
\end{document}